# PHÁT TRIỂN MẠNG CẢM BIẾN KHÔNG DÂY KẾT HỢP THIẾT BỊ BAY KHÔNG NGƯỜI LÁI PHỤC VỤ GIÁM SÁT CÂY NÔNG NGHIỆP

DEVELOPING SYSTEM OF WIRELESS SENSOR NETWORK AND UNMANED AERIAL VEHICLE FOR AGRICULTURE INSPECTION

Nguyễn Trường Sơn[2], Quách Công Hoàng[1,*], Đặng Thị Hương Giang[2,3], Vũ Minh Trung[2], Vương Quang Huy[2], Mai Anh Tuấn[1]

**TÓM TẮT**

Sản xuất nông nghiệp công nghệ cao là một xu hướng tất yếu tại Việt Nam. Đặc biệt với các vùng trồng cây nguyên liệu có đặc thù diện tích lớn, mạng cảm biến không dây đã và đang tỏ rõ vai trò quan trọng trong việc giúp tăng năng suất, giám sát sâu bệnh, làm giảm tác động của biến đổi khí hậu và giảm sức lao động trực tiếp của người canh tác. Bài báo này xây dựng một mô hình thử nghiệm giám sát vùng trồng cây nông nghiệp sử dụng phối hợp mạng cảm biến không dây LoRa và thiết bị bay không người lái nhằm thu thập các dữ liệu về điều kiện thời tiết, tình trạng đất, sức khoẻ cây trồng giúp người trồng đưa ra giải pháp phù hợp nhất về tưới tiêu, xử lý sâu bệnh, chăm bón với loại cây đang trồng. Hệ thống được phát triển và thực nghiệm ngoài hiện trường nhằm đánh giá một số tính năng cơ bản của hệ thống và chứng minh được tính ổn định, đáng tin cậy của dữ liệu thu được.

**Từ khóa:** Thiết bị bay không người lái, mạng cảm biến, nông nghiệp chính xác, LoRa.

**ABSTRACT**

Agricultural production using high technology is an inevitable trend in Vietnam. Especially for material crops which typically need large growing areas, wireless sensor networks has been clearly playing a significant role in increasing productivity, monitoring pests and diseases, mitigating the impact of climate change, and reducing the direct labor of cultivators. This paper constructs an experimental model of agricultural crop field monitoring using a combination of LoRa wireless sensor networks and unmanned aerial vehicles to collect data on conditions of weather and soil, plant health, which helps growers easily making right decisions on solutions for irrigation, pest treatment, and fertilization with the currently planted crops. The system has been developed and experimentized in the field to evaluate some basic features and justified the stability and reliability of the obtained data.

## 1. GIỚI THIỆU

Sử dụng các tiến bộ trong công nghệ thông tin và truyền thông để nâng cao năng suất nông nghiệp là một trong những giải pháp tiếp cận khả thi hiện nay. Cụ thể là trong những năm gần đây, sự ra đời của khái niệm Internet vạn vật (Internet of Things - IoT) và sự phổ biến nhanh chóng của các thiết bị bay không người lái (Unmanned Aerial Vehicle - UAV) kết hợp cùng công nghệ phân tích xử lý hình ảnh hứa hẹn các giải pháp nông nghiệp chính xác (Precision Agriculture - PA) để đương đầu với các thách thức được dự báo trong tương lai [1]. Một cách khái quát, nông nghiệp chính xác sử dụng các dịch vụ công nghệ thông tin để tổng hợp và xử lý dữ liệu thu thập được từ khu vực canh tác, sau đó đưa ra các đánh giá để hỗ trợ người nông dân quản lý mùa vụ một cách hiệu quả hơn [1-2].

Ứng dụng phổ biến nhất của nông nghiệp chính xác là đánh giá sinh trưởng cây trồng sử dụng công nghệ viễn thám và xử lý hình ảnh. Nguồn dữ liệu ảnh vệ tinh có giá thành cao đối với một người nông dân bình thường, hơn nữa độ phân giải và chất lượng hình ảnh chịu ảnh hưởng bởi điều kiện thời tiết [2]. Sử dụng UAV nhỏ trang bị các camera phổ chuyên dụng được xem như là lựa chọn kinh tế và an toàn nhất hiện tại. Các thông tin chỉ số sinh trưởng thực vật NDVI trên bản đồ xây dựng bởi ảnh chụp từ UAV có thể diễn dịch thành các dấu hiệu của một số vấn đề mà khu vực canh tác đang gặp phải như: sâu bệnh, thiếu nước và chất dinh dưỡng [1, 4-6] …

Mặc dù có nhiều ưu điểm hơn ảnh vệ tinh, việc thu thập dữ liệu ảnh bằng UAV cỡ nhỏ cũng bộc lộ một số hạn chế [3]. Thứ nhất, các UAV giá rẻ hiện nay chỉ có thể giám sát liên tục trong khoảng thời gian tối đa 30 phút với kết cấu dạng máy bay lên thẳng đa cánh quạt (multi-copter) và 90 phút với dạng máy bay cánh bằng. Thứ hai, người sử dụng cần phải lưu ý về thời gian và điều kiện thích hợp để vận hành UAV lấy mẫu ảnh phổ. Nếu không tuân thủ các bước này, dữ liệu thu thập sẽ dễ bị tác động bởi các điều kiện ánh sáng ngoại cảnh, gây khó khăn cho các bước phân tích dữ liệu về sau. Thứ ba, trong một khu vực canh tác tại Việt

Nam thường có các điều kiện thời tiết cục bộ khác nhau, điều này thường dẫn đến các rủi ro trong quá trình vận hành UAV mà người vận hành khó có thể lường trước nếu thiếu các công cụ hỗ trợ.

Để giải quyết vấn ba vấn đề trên, trong bài báo này chúng tôi đề xuất một mô hình giám sát nông nghiệp kết mạng cảm biến không dây kết hợp UAV. Trong hệ thống chúng tôi đề xuất, các nút mạng cảm biến không dây giá rẻ sử dụng sóng LoRa [14] được thiết lập trong các khu vực canh tác, giám sát ngày đêm các thông tin thời tiết, tưới tiêu, chiếu sáng… gửi về máy tính cơ sở dữ liệu để xử lý. Hệ thống này cho phép có được thông tin đầy đủ và liên tục về khu vực canh tác, đồng thời hỗ trợ các thiết bị UAV xác định được thời gian và địa điểm phù hợp để tiến hành khảo sát chụp ảnh. Chúng tôi còn tiến hành thử nghiệm mô hình đề xuất trên một khu vực canh tác tại Việt Nam, đánh giá hiệu quả và thảo luận cách thức thu thập và xử lý số liệu của hệ thống.

## 2. CÁC NGHIÊN CỨU LIÊN QUAN

Trong thập kỉ gần đây, ứng dụng UAV vào các nhiệm vụ giám sát hiện trường đã được nghiên cứu phát triển rộng rãi. Sự phổ biến của các UAV dân dụng cỡ nhỏ có giá thành thấp cùng sự phát triển của các thiết kế mở [3] đã tạo điều kiện thuận lợi cho các nghiên cứu này. Đặc biệt trong lĩnh vực giám sát nông nghiệp, UAV với cảm biến quang học phù hợp [1, 4-6] có thể thu thập những lớp thông tin về chỉ số sinh trưởng thực vật (NDVI, NDRE), kích thước của cây, nhiệt độ và độ ẩm khu vực canh tác... Kĩ thuật phân tích siêu phổ (hyperspectral) để phân tích sâu bệnh trên cây trồng đã bắt đầu được áp dụng trên UAV [1, 6] và hướng nghiên cứu này hứa hẹn trở thành xu thế trong tương lai không xa.

Mạng cảm biến không dây và rộng hơn là mô hình Internet vạn vật đang được nghiên cứu phát triển trong nhiều bài toán thực tế phục vụ đời sống con người mà nông nghiệp là một trong số đó [8-11]. Để quản lý một lượng lớn thiết bị cảm biến, giao thức truyền thông MQTT [12, 13] là một lựa chọn tiêu chuẩn cho phép giảm bớt sự quá tải các kết nối. Mạng cảm biến IoT sử dụng sóng LORA [8, 14] đã cho những kết quả khả quan trong việc tiết kiệm năng lượng và truyền tin khoảng cách xa, phù hợp với những khu vực xa đô thị. Để mở rộng phạm vi ứng dụng, tích hợp UAV trong mạng cảm biến IoT được đề xuất với nhiều vấn đề mở [11] như cấp phát tài nguyên, cơ chế bảo mật, phương thức phối hợp và thuật toán xử lý thông tin phù hợp [8, 11].

Điều kiện khí hậu và địa hình đa dạng của Việt Nam phù hợp với nhiều loại cây trồng có giá trị cao như chè, cà phê, hạt tiêu… Đây là nguồn tài nguyên thông tin dồi dào cho công nghệ thông tin và truyền thông tiếp cận phân tích và xây dựng các mô hình nông nghiệp chính xác. Tuy nhiên, diện tích lớn có khí hậu cục bộ không ổn định đòi hỏi phải giải quyết tốt các vấn đề an toàn bay và chất lượng dữ liệu thu [1, 3]. Mô hình kết hết hợp Internet vạn vật và UAV đề xuất bởi Yalin Liu [11] tỏ ra phù hợp với điều kiện thực tế tại Việt Nam.

## 3. THIẾT KẾ VÀ TRIỂN KHAI HỆ THỐNG

### 3.1. Tổng quan hệ thống

Mô hình hệ thống khảo sát của chúng tôi được xây dựng trên hai thành phần chính là mạng cảm biến không dây mặt đất sử dụng sóng LoRa và hệ thống UAV giám sát trên không. Trong đó, mạng cảm biến không dây LoRa có vai trò thu thập một cách liên tục các thông tin như nhiệt độ, độ ẩm, lượng mưa, tốc độ gió, áp suất, thành phần ánh sáng và không khí. Thiết bị bay có nhiệm vụ tự động thu thập ảnh phổ phản xạ cận hồng ngoại tại các khu vực được lên kế hoạch từ trước.

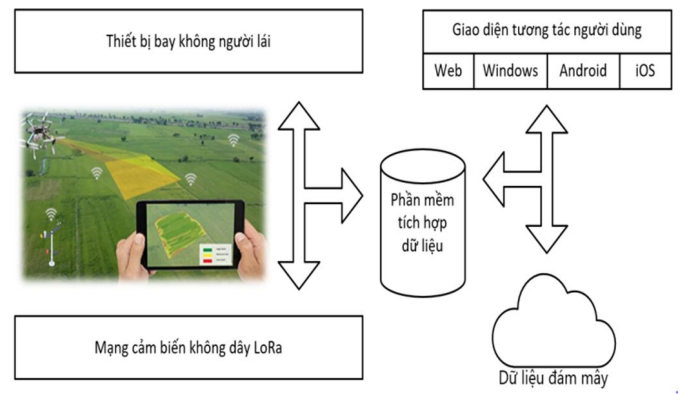

Hình 1. Lưu đồ tổng quát của hệ thống

Hệ thống phần mềm tích hợp thông tin cho phép duy trì kết nối thông tin liên tục giữa thiết bị bay và các nút cảm biến không dây hoạt động dưới mặt đất. Thiết kế tích hợp này được đề xuất với hai mục đích chính:

• Tích hợp thông tin về tình hình của khu vực canh tác, bao gồm: thông tin điều kiện môi trường từ các nút cảm biến cố định và thông tin hình ảnh đa mức phổ thu được từ UAV.

• Cảnh báo an toàn cho thiết bị bay về các điều kiện hiện tại của khu vực khảo sát, từ đó điều chỉnh kế hoạch bay khảo sát đánh giá cây trồng sao cho hiệu quả nhất.

Mô hình giám sát được phát triển dựa trên các thiết bị giá thành rẻ và các phần mềm mã nguồn mở. Điều này cho phép dễ dàng mở rộng và phát triển hệ thống, đáp ứng nhu cầu thực tế của nông nghiệp của Việt Nam trong hiện tại và tương lai.

### 3.2. Nút cảm biến không dây

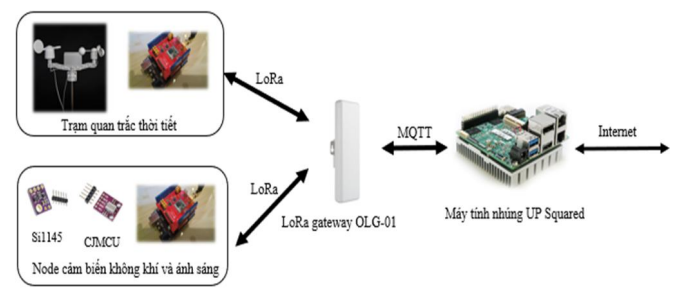

Hình 2. Lưu đồ quá trình truyền nhận thông tin từ cảm biến

Trong nghiên cứu này, chúng tôi thiết lập một mạng cảm biến không dây trên mặt đất sử dụng mạng truyền thông LoRa. Đây là một công nghệ truyền thông dữ liệu với mức năng lượng tiêu thụ thấp và khoảng cách truyền xa, được sử dụng rộng rãi trong các ứng dụng thu thập dữ liệu.

Dựa trên khảo sát về những loại dữ liệu môi trường cần được thu thập trong thực tế, chúng tôi xây dựng hệ thống các nút cảm biến không dây LoRa với hai nút cảm biến ngoài trời, một LoRa Gateway ngoài trời để tiếp nhận LoRa từ các nút cảm biến và một máy tính nhúng nhằm lưu trữ dữ liệu đồng thời gửi dữ liệu lên mạng internet. Các thành phần chính của hệ thống bao gồm:

• Máy tính nhúng là bộ xử lý trung tâm của hệ cảm biến LoRa và kết nối hệ cảm biến với cơ sở dữ liệu. Dữ liệu từ mạng cảm biến LoRa sẽ được tiến hành kiểm tra, sửa lỗi. Dữ liệu sau đó được lưu vào bộ nhớ của máy tính nhúng đồng thời được gửi tới máy chủ thông qua kết nối Internet. Việc này đảm bảo dữ liệu sẽ luôn được lưu trữ liên tục dù máy tính nhúng có bị mất kết nối internet.

• LoRa Gateway ngoài trời. Gateway này sẽ nhận dữ liệu từ các nút cảm biến thông qua mạng LoRa. Dữ liệu từ nút cảm biến sẽ được giải mã và được gửi cho máy tính nhúng dưới dạng các tin nhắn MQTT thông qua kết nối Ethernet.

• Các nút cảm biến có tính năng thu thập, xử lý dữ liệu của khu vực canh tác và truyền về trung tâm thông tin:

o Một trạm quan trắc thời tiết với các thông số đo đạc là nhiệt độ, độ ẩm không khí, vận tốc gió, hướng gió, áp suất không khí. Các thông tin này sẽ được sử dụng để đánh giá thời tiết cho sự phát triển của cây trồng cũng như phục vụ cho việc điều khiển quá trình bay của UAV.

o Cảm biến CJMCU dùng để đo các loại khí CO, $NH_3$, $NO_2$ trong không khí. Các thông số này ảnh hưởng rất nhiều tới quá trình sinh trưởng của cây trồng.

o Cảm biến ánh sáng Si1145. Thông số này sẽ được sử dụng để đánh giá điều kiện môi trường đồng thời là một tham số để hiệu chỉnh hình ảnh trong quá trình xây dựng bản đồ ảnh của UAV.

Nhằm khảo sát độ tin cậy và khả năng hoạt động của hệ thống, chúng tôi đã tiến hành lắp đặt hệ thống ngoài trời và tiến hành đo đạc khả năng truyền thông của các nút cảm biến. Theo thông số của nhà sản xuất, thiết bị LoRa chúng tôi sử dụng có hệ số lan tỏa (Spreading Factor) là 7, băng thông tín hiệu truyền là 125kHz, tốc độ mã hóa là 4/5. Trong thực nghiệm này, chúng tôi cấu hình thiết bị ở tần số 923MHz và công suất phát của mạch đo đạc được là 20dBm. Chúng tôi sử dụng ba an-ten có công suất phát khác nhau là 3dBi, 10dBi và 18dBi. Kết quả được đo bởi thiết bị RF Explorer 915M V2.0 [15].

Với điều kiện thực nghiệm trên mặt phẳng trải dài không có vật cản xung quanh, kết quả thu được chứng tỏ rằng khi sử dụng an-ten có công suất phát 18 dBi với khoảng cách 450m, tín hiệu LoRa vẫn có thể nhận được bởi các thiết bị thu sóng LoRa với độ nhạy lớn hơn -100dBm. Thiết bị thu sóng mà chúng tôi sử dụng là LoRa Gateway OLG-01 hoàn toàn đáp ứng được yêu cầu này với độ nhạy lên tới -120dBm.

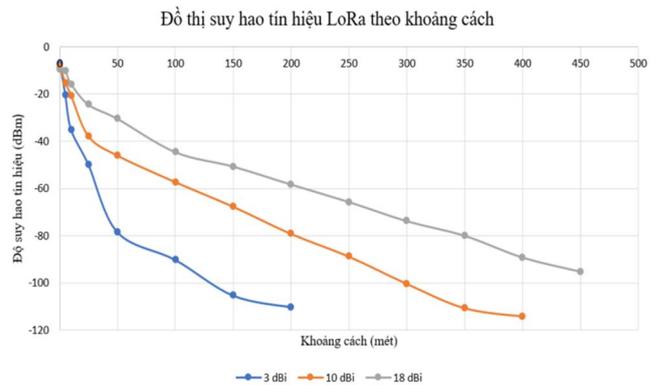

Hình 3. Đồ thị suy hao tín hiệu LoRa theo khoảng cách

### 3.3. Thiết bị bay giám sát nông nghiệp

Thiết bị bay chụp ảnh nông nghiệp của hệ thống giám sát được chúng tôi phát triển dựa trên nghiên cứu [7] với phần lớn là các linh kiện giá rẻ sẵn có tại Việt Nam. Chúng tôi đã có một số thay đổi về thiết kế cơ khí để thiết bị bay có thể mang thêm 01 camera chụp ảnh phổ chuyên dụng cho cây trồng.

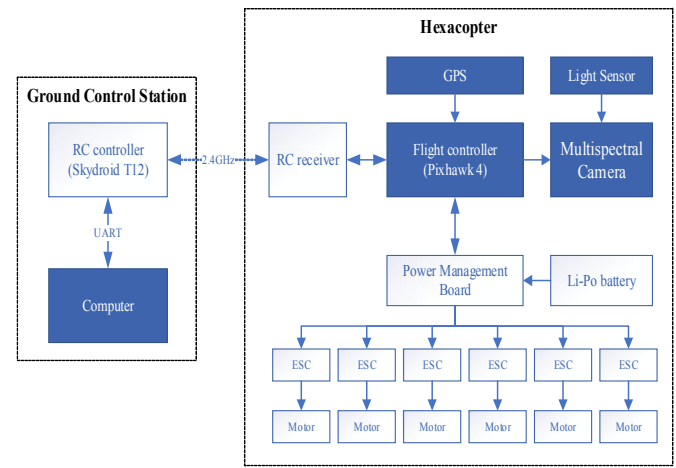

Hình 4. Sơ đồ phần cứng của UAV

Trong hình 4 mô tả kết nối của thiết bị bay mà chúng tôi đã sử dụng, các thiết bị chính của UAV bao gồm:

• Bộ điều khiển bay Pixhawk 4: kết nối với GPS cho phép điều hướng thiết bị bay đến một vị trí xác định trên bản đồ.

• Hệ camera chụp ảnh đa phổ Micasense RedEdge-M: có vai trò nhận tín hiệu điều khiển chụp ảnh từ bộ điều khiển bay. Bên cạnh đó, do tích hợp 01 cảm biến ánh sáng, hệ camera này có thể hiệu chỉnh màu sắc của ảnh khi điều kiện chiếu sáng bị thay đổi. (citation)

• Bộ điều khiển từ xa Skydroid T12: có nhiệm vụ gửi các tín hiệu điều khiển hay quỹ đạo bay chụp ảnh mong muốn tới bộ điều khiển bay Pixhawk 4 bằng sóng mang 2.4GHz. Ngoài ra tay cầm cũng cho phép nhận các tín hiệu trạng thái do thiết bị bay gửi về như tọa độ GPS, thông tin IMU, cảm biến gió…

• Máy tính với kết nối mạng: thiết lập các quỹ đạo chụp ảnh mong muốn để gửi lên bộ điều khiển bay Pixhawk 4 thông qua bộ điều khiển từ xa.

Để thiết bị bay mà chúng tôi tích hợp có trọng tải cất cánh 2,5kg, tốc độ bay lên tới 36km/h. Thời gian bay khảo sát thực tế là 12 phút đối với pin lipo 15V 5200mAh và 17 phút đối với pin lipo 15V 10000mAh. Tuy nhiên thời gian này còn phụ thuộc nhiều vào điều kiện thời tiết tại khu vực hoạt động, tiêu biểu như hướng gió và khí áp. Thực tế là khí áp ảnh hưởng tới lực nâng của cánh quạt trên thiết bị bay, khí áp thấp khiến cánh quạt phải quay nhanh hơn, kéo theo thời lượng pin giảm xuống. Đối với quỹ đạo bay, việc di chuyển ngược gió trong thời gian dài cũng khiến động cơ tiêu tốn nhiều điện năng. Từ hai thí dụ trên cho chúng ta thấy phần mềm điều khiển lập quỹ đạo bay cần biết thêm các thông tin về trạng thái môi trường khảo sát mới đủ cơ sở để ước lượng quỹ đạo tối ưu [3].

### 3.4. Phần mềm hệ thống tích hợp mạng cảm biến và thiết bị bay

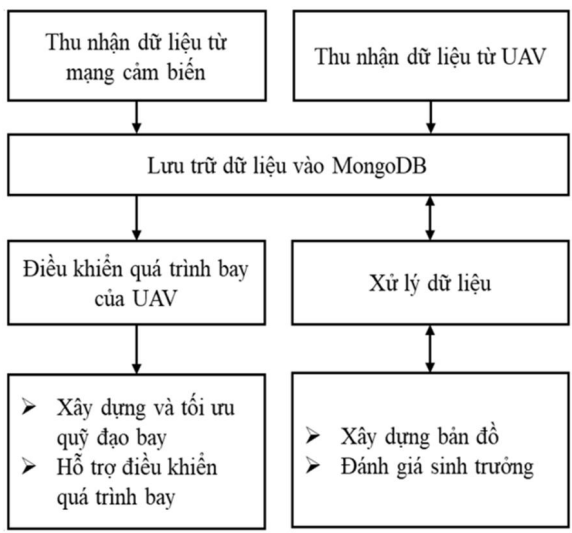

Hình 5. Tổng quan phần mềm hệ thống tích hợp mạng cảm biến và thiết bị bay

Sau khi thiết kế và thực nghiệm thành công hai thành phần mạng cảm biến và thiết bị bay, chúng tôi xây dựng một phần mềm để tích hợp hai hệ thống này. Phần mềm này có vai trò như bộ xử lý trung tâm cho toàn bộ hệ thống. Phần mềm được xây dựng để hệ thống có thể hoạt động với sự can thiệp ít nhất của con người, từ đó người nông dân có thể dễ dàng sử dụng hệ thống chỉ với những bước cài đặt cơ bản. Các tác vụ chính của phần mềm như sau:

• Thu thập dữ liệu cảm biến môi trường: mô-đun có vai trò kết nối với LoRa Gateway thông quan giao thức MQTT để tiếp nhận từ liệu từ cảm biến. Sau khi tiếp nhận mô-đun này cũng sẽ kiểm tra tính xác thực của dữ liệu để chuyển tới mô-đun lưu trữ.

• Thu thập dữ liệu từ cảm biến của UAV: mô-đun có nhiệm vụ tiếp nhận dữ liệu từ UAV bằng giao thức MAVLink. Thông qua kết nối với Skydroid T12 mô-đun này sẽ tiếp nhận các thông tin từ UAV gửi về trong quá trình bay và gửi tới mô-đun lưu trữ.

• Lưu trữ dữ liệu vào MongoDB: mô-đun này sẽ tiếp nhận dữ liệu từ 2 mô-đun thu thập dữ liệu bên trên, sau đó chuyển thành dạng dữ liệu chuẩn và lưu trữ vào MongoDB. Ngoài ra mô-đun cũng sẽ cung cấp những chức năng cơ bản để thống kê, hiển thị dữ liệu dưới dạng đồ thị.

• Điều khiển quá trình bay của UAV: mô-đun có nhiệm vụ điều tiết quá trình bay của UAV. Mô-đun này sẽ sử dụng dữ liệu về môi trường được lưu trữ trong MongoDB, thống kê lại để xây dựng các nhiệm vụ bay cho UAV cũng như các công việc khác như tối ưu quỹ đạo, tốc độ di chuyển…

• Xử lý dữ liệu: Mô-đun này sẽ sử dụng các dữ liệu được lưu trữ trong cơ sở dữ liệu để làm các nhiệm vụ như xây dựng bản đồ ảnh phổ khả kiến RGB và bản đồ chỉ số sinh trưởng cây trồng như NDVI (Normalized Difference Vegetation Index) hay NDRE (Normalized Difference Red Edge Index). Bản đồ này kết hợp với thông tin tổng hợp về điều kiện tưới tiêu và chiếu sáng là cơ sở chính để phát triển các dịch vụ đánh giá sinh trưởng cây trồng trong tương lai.

Các bộ thư viện mã nguồn mở được sử dụng trong thiết kế phần mềm của hệ thống bao gồm:

• Thư viện truyền nhận thông tin Eclipse Mosquitto: Là thư viện hỗ trợ giao thức MQTT cho mô-đun thu thập dữ liệu cảm biến môi trường.

• Thư viện MAVLink: Là thư viện hỗ trợ giao thứ MAVLink cho mô-đun thu thập dữ liệu của UAV.

• Cơ sở dữ liệu MongoDB: Có nhiệm vụ chính là xây dựng một cơ sở dữ liệu nhằm lưu trữ dữ liệu từ cảm biến và UAV một cách khoa học cho quá trình xử lý dữ liệu sau này.

• Phần mềm giao diện Qt: Có nhiệm vụ xây dựng giao diện giao tiếp với người dùng và đồ thị hóa dữ liệu cũng như hiển thị dữ liệu mà người dùng yêu cầu.

Hệ thống tích hợp cảm biến không dây cho phép can thiệp vào quá trình bay giám sát nông nghiệp theo quy tắc sau:

- Cảnh báo không đảm bảo khả năng lấy mẫu ảnh phổ cận hồng ngoại trong một số trường hợp: khi trời nhiều mây, góc chiếu của mặt trời thấp dưới 45 độ.

- Không cho thiết bị bay cất cánh trong điều kiện mưa, hoặc vận tốc gió tại khu vực giám sát có vận tốc trung bình trên 10m/s.

- Tối ưu quỹ đạo bay trong điều kiện gió từ 3 đến 10m/s.

Quy tắc sinh quỹ đạo bay trong điều kiện trên được chúng tôi xây dựng nhằm hạn chế tối đa thời gian bay ngược gió của UAV. Quỹ đạo được thành lập cụ thể theo các bước như sau:

- Khu vực giám sát được định nghĩa bởi một đa giác A, trần bay được giữ nguyên trong suốt quá trình bay.

- Định nghĩa cặp véc tơ trực giao $\vec{i}$ và $\vec{j}$, trong đó $\vec{j}$ có phương song song với hướng gió. Độ lớn của hai véc tơ phụ thuộc vào thị trường của camera và trần bay của UAV.

- Thiết bị bay sẽ di chuyển chụp ảnh tịnh tiến theo phương $\vec{i}$ trong suốt quá trình giám sát và chỉ thay đổi $\vec{j}$ khi điểm chụp ảnh nằm ngoài đa giác A [16].

## 4. KHẢ NĂNG GIÁM SÁT HÌNH ẢNH TRONG VÙNG CÂY NGUYÊN LIỆU

Chúng tôi thử nghiệm hệ thống trong một khu vực canh tác có diện tích 1 héc ta tại khu vực đồng bằng Bắc Bộ Việt Nam. Các nút cảm biến không dây được thiết lập trên mặt ruộng, cách Lora gateway và máy tính trung tâm 50 mét. Tại thời điểm đo đạc, nhiệt độ môi trường là 33⁰C, độ ẩm 40%, tốc độ gió trung bình 8,5km/h. Lúc thiết bị bay cất cánh, tức 14 giờ 30, ánh sáng tử ngoại là 7,9UV index còn vùng khả kiến và hồng ngoại dao động trong khoảng 80 - 90 klx.

Thiết bị chụp ảnh đa phổ chuyên dụng được chúng tôi sử dụng là Micasense RedEdge-M. Thiết bị này có khả năng chụp không ảnh đồng thời 05 dải sóng:

- Xanh da trời (Blue): 475nm ± 20nm
- Xanh lá cây (Green): 560nm ± 20nm
- Đỏ (Red): 668nm ± 10nm
- Cận hồng ngoại (Near IR): 840nm ± 40nm
- Biên đỏ (Red Edge): 717nm ± 10nm

Như chúng tôi đề cập ở phần trước, qua giao thức MavLink, hệ thống mặt đất đề xuất lịch trình bay gửi lên bộ điều khiển bay Pixhawk 4. Trong quá trình thực thi lịch trình bay liên tục cập nhật tọa độ định vị GPS, bộ điều khiển sẽ xác định vị trí phù hợp để chụp hình và gửi lệnh điều khiển xuống thiết bị Micasense RedEdge-M. Sau khi UAV kết thúc quá trình khảo sát, dữ liệu ảnh đa phổ được lưu trong thẻ nhớ được đưa vào các phần mềm phân tích ảnh chuyên dụng.

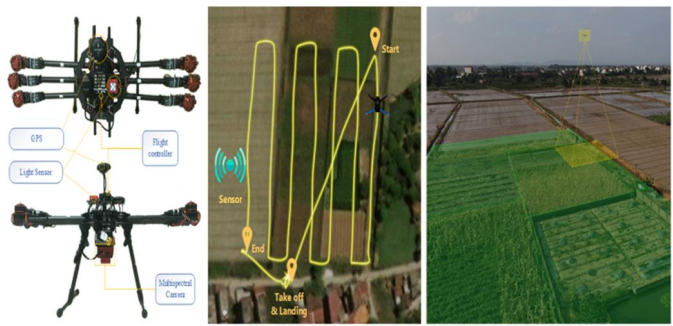

Hình 6. Bố trí thực nghiệm với hệ thống chụp ảnh phổ bằng UAV

Để đánh giá kết quả thu được từ quá trình giám sát, chúng tôi sử dụng một số phần mềm tiên tiến nhất là Pix4Dmapper, Pix4Dfields. Đây là các phần mềm thương mại được đánh giá cao, mỗi phần mềm được thiết kế có ưu nhược điểm riêng. Các phần mềm này cùng được thực thi trên máy trạm với CPU Intel Xeon 2620v4 2.2GHz dùng hệ điều hành Windows 10. Để các đánh giá được khách quan, chúng tôi còn bổ sung thêm một số bộ dữ liệu mở được thu thập từ UAV dạng máy bay cánh bằng.

Bảng 1 cho thấy Pix4Dmapper mặc dù cho chất lượng bản đồ chính xác nhất hiện nay tuy nhiên cũng tiêu tốn nhiều thời gian tính toán, so với Pix4Dfields chênh lệch khoảng 10 lần. Chất lượng bản đồ của Pix4Dfields trong hình 8 có một số điểm không tốt bằng Pix4Dmapper, nguyên nhân là camera chụp bị nghiêng do ảnh hưởng của gió giật và UAV không được trang bị chống rung.

Bảng 1. So sánh thời gian xử lý đối với dữ liệu ảnh đa phổ thu thập được

|  | Điều kiện khảo sát | | | Số lượng ảnh | Phân giải | Thời gian xử lý | |
| --- | --- | --- | --- | --- | --- | --- | --- |
|  | Trần bay | Diện tích | Số lớp phổ |  |  | Pix4D Mapper | Pix4D Fields |
| Dữ liệu thu thập thực địa (Việt Nam) | 50 mét | 1 ha | 5 (Red, Green, Blue, Near IR, Red Edge) | 785 | 1280 x 960 | 31 phút | 4 phút |
| Dữ liệu mở (internet) | 120 mét | 100 ha | 4 (Red, Green, Near IR, Red Edge) | 5260 | 1280 x 960 | 179 phút | 18 phút |

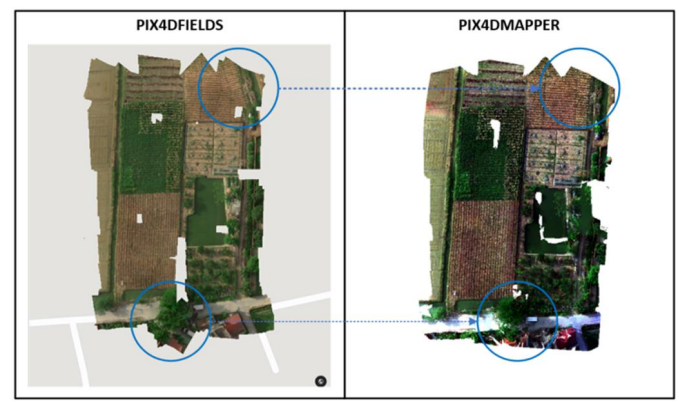

Hình 7. So sánh chất lượng bản đồ ảnh xây dựng bởi phần mềm Pix4Dfields và Pix4Dmapper

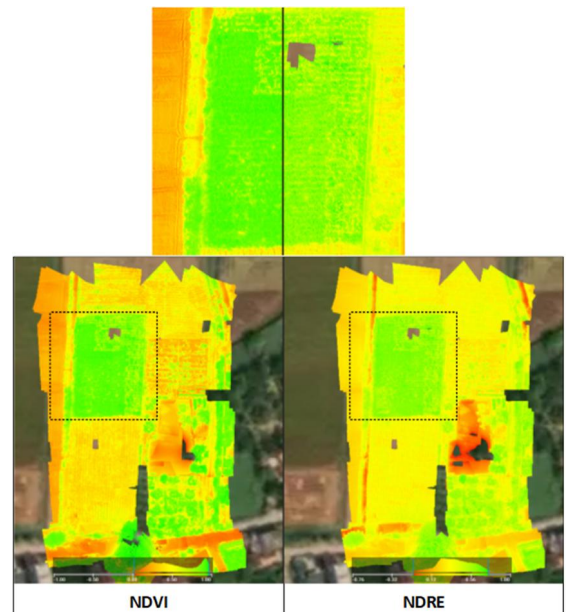

Hình 8. So sánh chỉ số sinh trưởng cây trồng NDVI và NDRE

Từ dữ liệu bản đồ chụp bởi camera đa phổ, chúng tôi sử dụng hai chỉ số phổ biến để đánh giá sinh trưởng cây trồng là NDVI và NDRE [1]. Chỉ số NDVI được đánh giá dựa trên tương phản của bức xạ phổ hồng ngoại (NIR) với phổ màu đỏ (Red) bởi công thức sau:

$$\text{NDVI} = \frac{(\text{NIR} - \text{Red})}{(\text{NIR} + \text{Red})} \quad (1)$$

Khác với NDVI, chỉ số NDRE dựa trên tương phản với phổ cận đỏ (RE) 715nm:

$$\text{NDRE} = \frac{(\text{NIR} - \text{RE})}{(\text{NIR} + \text{RE})} \quad (2)$$

Từ hình 8 và 9 chúng ta có thể thấy sự khác biệt của hai chỉ số trên tại các khu vực mà cây trồng đang sinh trưởng. Tương phản của NDRE rõ rệt hơn NDVI do NDRE nhạy cảm với hàm lượng diệp lục, sự thay đổi của diện tích lá, ảnh hưởng bởi đất nền [1]. Vậy nên trong giai đoạn giữa và cuối mùa vụ, việc khảo sát dải phổ cận đỏ để xác định chỉ số NDRE sẽ đem lại nhiều thông tin chi tiết hơn.

**5. KẾT LUẬN**

Trong bài báo này, chúng tôi đã mô tả một mô hình hệ thống giám sát vùng nông nghiệp sử dụng kết hợp mạng cảm biến không dây và thiết bị bay không người lái. Hệ thống mạng cảm biến không dây và thiết bị bay được tính toán và xây dựng dựa trên các trang thiết bị giá thành thấp, phù hợp với điều kiện nhu cầu thực tế tại Việt Nam. Các thử nghiệm trên điều kiện thực địa cho thấy khả năng giám sát sinh trưởng cây trồng của hệ thống với một khu vực canh tác tiêu biểu tại Việt Nam. Trong tương lai, việc sử dụng các phần cứng và phần mềm thiết kế mở cho phép hệ thống có thể dễ dàng phát triển và hoàn thiện các tính năng tự động chụp ảnh và xây dựng bản đồ. Hệ thống hứa hẹn cho phép xây dựng cơ sở dữ liệu số hóa nông nghiệp tại đồng ruộng Việt Nam với quy mô dữ liệu lớn và giá thành thấp; đây là tiền đề cho việc áp dụng công nghệ học máy cho bài toán phân tích sinh trưởng cây trồng tại Việt Nam trong tương lai.

**AUTHORS INFORMATION**

**Nguyen Truong Son[2], Quach Cong Hoang[1], Dang Thi Huong Giang[2,3], Vu Minh Trung[2], Vuong Quang Huy[2], Mai Anh Tuan[1]**

[1]Nacentech Technology and Business Incubator Center, National Center for Technological Progress

[2]VNU University of Engineering and Technology

[3]University of Economics - Technology for Industries